\documentclass{emulateapj}

\def\til{\ensuremath{\sim\,}}

\def\chisq{\ensuremath{\chi^2}}
\def\rchisq{\ensuremath{\chi_{\nu}^{2}}}

\newcommand{\tim}[1]{\ensuremath{\times 10^{#1}}}

\def\etal{et al.\ }

\def\mekal{{\sc mekal}}
\def\mwd{\ensuremath{M_{\rm WD}}}
\def\msol{\ensuremath{M_\odot}}
\def\xmm{\emph{XMM}}
\def\xmmn{\emph{XMM-Newton}}
\def\cms{\ensuremath{$cm$^{-2}}}

\def\mdot{\ensuremath{\dot{m}}}

\def\dcs{\ensuremath{\Delta\chisq_{\rm system}}}

\shorttitle{Soft X-rays from IPs}
\shortauthors{Evans \&\ Hellier}

\begin{document}

\author{P.A. Evans\altaffilmark{1} and Coel Hellier}
\affil{Astrophysics Group, Keele University, Staffordshire, ST5 5BG, UK}
\altaffiltext{1}{Current address: Department of Physics and Astronomy,
University of Leicester, Leicester, LE1 7RH, UK}

\title{Why do some intermediate polars show soft X-ray emission? A survey of
\xmmn\/ spectra.}

\begin{abstract} 
We make a systematic analysis of the \xmmn\ X-ray spectra of intermediate polars
(IPs) and find that, contrary to the traditional picture, most show a soft
blackbody component.  We compare the results with those from AM~Her stars and
deduce that the blackbody emission arises from reprocessing of hard X-rays,
rather than from the blobby accretion sometimes seen in AM Hers. Whether an IP
shows a blackbody component appears to depend primarily on geometric factors: a
blackbody is not seen in those that have accretion footprints that are always
obscured by accretion curtains or are only visible when foreshortened on the
white-dwarf limb. Thus we argue against previous suggestions that the blackbody
emission characterises a separate sub-group of IPs which are more akin to AM
Hers, and develop a unified picture of the blackbody emission in these stars. 
\end{abstract}

\keywords{accretion, accretion discs -- novae, cataclysmic variables --
X-rays:
binaries.}

\section{Introduction}
\label{sec:intro}

Intermediate polars (IPs) -- interacting binaries with a magnetic white-dwarf
primary -- have traditionally been noted for their hard X-ray emission. This
arises as the magnetic field of the white dwarf disrupts the accretion disc and
channels material towards the magnetic polecaps. This material forms stand-off
shocks, below which it cools via free-free interactions, producing hard X-rays.
However, a growing number of systems have been shown to emit a distinct
blackbody component in softer X-rays (e.g.\
\citet{Mason92,Haberl94,deMartino04}), reminiscent of the soft component
prominent in the X-ray spectra of many AM Her stars (also known as polars).
These systems are similar to IPs but the white-dwarf has a magnetic field strong
enough to prevent an accretion disc from forming at all. In these systems, the
soft blackbody component is though to arise from a heated polecap surrounding
the accretion column (See \citet{Warner95,Hellier01} for a review of these
objects).

Currently it is unclear why the blackbody component is seen in some IPs and not
others.  \citet{Haberl95} suggested that there are two distinct classes of IP,
with the `soft' systems being evolutionary progenitors of polars. They argued
that the `hard IPs' may have larger and cooler polecaps, pushing the soft
emission into the EUV and explaining the difference in the spectra.

We present here a study of \xmmn\ X-ray data of 12 IPs, aimed at discovering why
some IPs show a blackbody component while others don't. Our method is similar to
that of \citet{Ramsay04} (hereafter RC04) who analysed the \xmmn\ data of
twenty-one polars, which enables us to compare the  IPs with the polars.

\section{Observations and data analysis}
\label{sec:obs}

The \xmmn\/\ observatory \citep{Jansen01} was launched in 1999, and we have
obtained observations of twelve IPs from the public archive. We analyzed the
data from the EPIC-MOS and pn instruments \citep{Turner01, Struder01} which
provide high-throughput, medium-resolution spectroscopy across the 0.2--12 keV
energy range. The higher resolution RGS instruments \citep{denHerder01} have
only 20 per cent of the effective area of the MOS cameras and the data are not
used here.

A summary of the observations used is given in Table~\ref{tab:obs}. We re-ran
the pipeline processing for these observations using {\sc xmm-sas} v7.0.0. The
observations of GK~Per, NY Lup and V2400~Oph suffered from pile-up, and thus
only the wings of the PSF were included in the source extraction. The MOS-1
observation of EX~Hya was so badly piled up that we excluded it from our
analysis.

\begin{deluxetable}{lccl}
\tablecaption{The \xmm\/\ observations of IPs analysed in this paper.}
\tablehead{\colhead{Object} & \colhead{ObsID} & \colhead{Date} &
\colhead{References}}
\startdata
AO~Psc     &  0009650101 & 2001-06-09 & 1,2\\
EX~Hya     &  0111020101 & 2000-07-01 & 1,2 \\
           &  0111020201 & 2000-07-01 & 1,2 \\
FO~Aqr     &  0009650201 & 2001-05-12 & 1,2,3 \\
GK~Per     &  0154550101 & 2002-03-09 & 2,4\\
HT~Cam     &  0144840101 & 2003-03-24 & 2,5,6 \\
PQ~Gem     &  0109510301 & 2002-10-08 & 2,7 \\
NY~Lup     & 0105460301 & 2000-09-07 & 2,8 \\
UU~Col     &  0201290201 & 2004-08-21 & 9 \\
V1223~Sgr  &  0145050101 & 2003-04-13 & 1,2\\
V405~Aur   &  0111180401 & 2001-10-05 & 2,10 \\
V2400~Oph  &  0105460601 & 2001-08-30 & 2 \\
WX~Pyx     &  0149160201 & 2003-05-20 & 11 \\
\enddata
\tablerefs{(1) Cropper \etal(2002), (2) Evans \&\ Hellier (2005b), (3) Evans
\etal(2004), (4) Vrielmann \etal(2005) (5) de~Martino \etal(2005), (6)
Evans \&\
Hellier (2005a), (7) Evans, Hellier \&\ Ramsay (2006), (8) Haberl \etal
(2002),
(9) de~Martino \etal(2006) (10) Evans \&\ Hellier (2004), (11) Schlegel
(2005).}
\label{tab:obs}
\end{deluxetable}

RC04 used only the EPIC-pn data as it was better calibrated than the EPIC-MOS
data at soft energies. Using the better calibrations of {\sc xmm-sas v.}7 we
extracted spectra from all three EPIC instruments. Response matrices were
created for each spectrum, using the {\sc xmm-sas} rmfgen and arfgen tasks. We
then modelled the spectra using {\sc xspec~v}11. For each star, all model
parameters were tied between the EPIC instruments, except for the normalisation
which we allowed to vary in order to combat the effects of cross-calibration
uncertainties.

Although IP spectra can vary considerably over the spin cycle, for the majority
of the systems in this paper, we do not have enough geometric information to
identify phase regions when the hard/soft components are best presented to us (as
RC04 did), so we extracted spectra covering the entire observation. Note that
the results of our spectroscopy are thus weighted averages from across the spin
cycle; this was taken into account when interpreting our results .

To reproduce the hard component we used the stratified accretion column model of
\citet{Cropper99}. This models the spectrum in terms of the white dwarf mass
(\mwd) and specific accretion rate (i.e.\ accretion rate per unit area, \mdot),
from which it calculates the temperature and density profile of the column. This
is then divided into 100 bins, evenly distributed in velocity space, each bin
emitting as an optically thin plasma (a \mekal). To the stratified column model,
we added narrow Gaussians for the 6.4-keV iron fluorescence line and the 0.547
keV Oxygen {\sc vii} photoionisation line where necessary. We then applied to
this emission a simple photoelectric absorber. For most systems this did not
give an acceptable fit, so we added either one or two  partial-covering
absorbers as necessary.

Next, we added a blackbody component to the models. Since absorption at the
densities of the partial-covering components (typically \til10$^{23}$ \cms) will
completely smother any soft X-ray emission and thus be redundant with model
normalisation, the blackbody component was absorbed only by the simple
absorption, which was of order 10$^{19}$--10$^{21}$ \cms.

For some systems the addition of a blackbody did not improve the fit. For these
systems we manually raised the blackbody normalisation until it significantly
reduced the fit quality, thus finding an upper limit. Since this will be
temperature dependent, we did this for blackbody temperatures of 40, 60 and 80
eV.

We quote, in Table~\ref{tab:tests}, the f-test statistic to judge the the
significance of adding the blackbody component.  However, this test will produce
false positives in the presence of calibration systematics.  We have thus
estimated the systematics by fitting a model optimised for the MOS data to the
pn data (allowing only the normalisation to change) and recording the change in
\chisq (=\dcs). We claim the presence of a blackbody only if it improves the
\chisq\ by more than \dcs. This method is more conservative than using the
f-test alone.  We include this estimate of the systematics in all the errors
quoted in this paper.

Details of the fits are given in Table~\ref{tab:comps}. The \mdot\ was
unconstrained for every system, so is not given. We do not quote errors on the
partial-covering absorbers as they do not affect the softness ratio. The ratio
is sensitive, however, to the metal abundance in the column, as there is a
forest of iron $L$ lines in the 0.5--1.2 keV range, affecting the model fit at
the soft end.

%%%%%%%%%%%%%%%% F-tests and related animals %%%%%%%%%%%%%%%%

\begin{deluxetable*}{cccccc}
\tablecaption{Fit statistics for each star with and without a blackbody.}
\tablehead{\colhead{Star} &
\colhead{\chisq\ (dof)} & \colhead{\chisq\ (dof) (with
bb)} & \colhead{f-test} & \colhead{$\dcs$} & \colhead{$\chisq_{\rm
nobb}-\chisq_{\rm bb}$  } \\ & \colhead{(No bb)} & \colhead{(with bb)}
}
\startdata
AO~Psc     &  3372.74 (2888)     &  3772.71 (2884)  &  0.93       &  700
    & 0.03  \\
FO~Aqr     &  1462  (1864)       &  1444 (1860)     &  1.6\tim{-4}&  168
    & 18 \\
HT~Cam     &  1325 (1273)        &  1310 (1269)     &  4.1\tim{-3}&  79
     & 16 \\
V1223 Sgr  &  3840.8 (3128)      &  3840.6 (3124)   &  0.99       &  143
    & 0.2 \\
EX~Hya     &  14045 (4876)       &  10158 (4873)    & $<10^{-99}$ &
1284    & 3887 \\
GK Per     &  17040 (4079)       &  5024 (4075)     & $<10^{-99}$ &  84
     & 12016 \\
NY~Lup     &  902 (699)          &  669 (695)       &  8\tim{-14} &  40
     & 233 \\
PQ~Gem     &  20435 (2439)       &  2940 (2435)     & $<10^{-99}$ &  144
    & 17495 \\
UU~Col     &  1676 (817)         &  910 (813)       & $<10^{-99}$ &  76
     & 766 \\
V2400~Oph  &  1019 (1003)        &  953 (999)       &  9\tim{-14} &  50
     & 66.15\\
V405 Aur   &  17626 (997)        &  1146 (994)      & $<10^{-99}$ &  *
     & 16480 \\
WX~Pyx     &  594 (477)          &  495 (473)       &  8\tim{-18} &  52
     & 99 \\
\enddata
\tablecomments{ The f-test gives the probability that no blackbody is
present,
making no allowance for systematics. The \dcs\ is the change in \chisq\ in
fitting the same model to the MOS and pn cameras, thus giving an
estimate of the
systematic errors. The last column is the improvement in \chisq\ when a
blackbody is added. We consider this significant if it exceeds \dcs.
$^*$ There was no pn data for V405~Aur, so \dcs\ was not estimated. }
\label{tab:tests}
\end{deluxetable*}

%%%%%%%%%%%%%%%%%%%% SPECTRAL COMPONENTS %%%%%%%%%%%%%%%%%

\begin{deluxetable*}{ccccccc}
\tabletypesize{\scriptsize}
\tablecaption{Spectral components used in the fitted models.}
\tablehead{\colhead{Star} & \colhead{wabs $n_H$} & \colhead{blackbody kT} &
\colhead{Part Abs (1)} & \colhead{Part Abs (2)} & \colhead{\mwd}  &
\colhead{Abundance} \\
 & \colhead{(10$^{20}$\cms)} & \colhead{(eV)} & \colhead{($n_H$, Cv Frc)} &
\colhead{($n_H$, Cv Frc)} & \colhead{(\msol)} & \colhead{(solar)}}
\startdata
V405 Aur   &  3.46 (+0.41, $-$0.31) & 64.78 (+0.81, $-$1.11 ) & 17, 0.49
       & 3.0, 0.63          &  0.40 (+0.05, $-$0.06)  & 0.069 (+0.024,
$-$0.021) \\
GK Per     &  23.3 (+2.0, $-$1.9)   & 62 ($\pm2$)             & 23, 0.74
       & 4.7, 0.45          &  0.92 (+0.39, $-$0.13)  & 0.21 (+0.14,
$-$0.07) \\
NY~Lup     &  7.8  ($\pm3.9$)       & 104 (+21, $-$23)        & 14, 0.49
       & 0.38, 0.71         &  0.96 (+0.40, $-$0.55)  & 0.68 (+0.51,
$-$0.59) \\
V2400~Oph  &  7.0 (+2.9, $-$4.9)    & 117 (+33,  $-$44)       & 11, 0.52
       & 0.61, 0.53         &  0.69 (+0.06, $-$0.24)  & 0.33 (+0.12,
$-$0.10) \\
PQ~Gem     &  0 (+0.30)    & 47.6 (+2.9, $-$1.4)     & 42, 0.60
       & 3.4, 0.56          &  0.70 (+0.16, $-$0.14)  & $<0.08$ \\
EX~Hya     &  9.76 (+2.2,  $-$0.86) & 31.0 (+1.3, -2.4)       & 75, 0.35
       & 4.0, 0.29          &  0.449 (+0.005, $-$0.013) & 0.514 (+0.01,
$-$0.0029) \\
UU~Col     &  0 (+0.59)           & 73 (+20, $-$9)          & 10, 0.34
       &                    &  1.23 (+0.17, $-$0.29)  & 0.66 (+1.0,
-0.62) \\
WX~Pyx     &  8.4 (+3.8,  $-$2.9)   & 82 (+11, $-$15)         &
        &                    &  1.4 (+0, $-$0.09)      & $<2.87$  \\
FO~Aqr     &  0 (+2.1)              &                         & 21, 0.80
       & 6.4,  0.98         &  1.19 (+0.11, $-$0.31)  & 0.31 (+0.20,
$-$0.23) \\
AO~Psc     &  3.89 (+0.69, $-$1.44) &                         & 14, 0.62
       & 1.8,  0.75         &  0.594 (+0.13, $-$0.040)& 0.362 (+0.20,
$-$0.064) \\
HT~Cam     &  3.86 (+0.81, -0.88)   &                         &
        &                    &  0.687 (+0.094, $-$0.061)& 0.52 (+0.24,
$-$0.11) \\
V1223 Sgr  &  1.03 (+0.36, $-$0.52) &                         & 13, 0.46
       & 1.3, 0.63          &  1.046 (+0.049, $-$0.012) & 0.398 (+0.090,
-0.049) \\
\enddata
\tablecomments{The column density of the partial absorption is given in
units of
$10^{22}\ \cms$. Errors are quoted to the same power of ten as the
corresponding
parameter.}
\label{tab:comps}
\end{deluxetable*}

For all twelve systems we then calculated the flux from the hard and soft
components. Following RC04 we defined the softness ratio as \mbox{$ F_{s}/
4F_{h}$}, where $F_s$ and $F_h$ are the fluxes of the soft and hard components
respectively. The factor of four arises because the hard component is optically
thin and thus radiates isotropically, whereas the hard component is optically
thick. Where the blackbody-emitting region is seen foreshortened, the observed
ratio will be an underestimate.

\begin{figure*}
\plotone{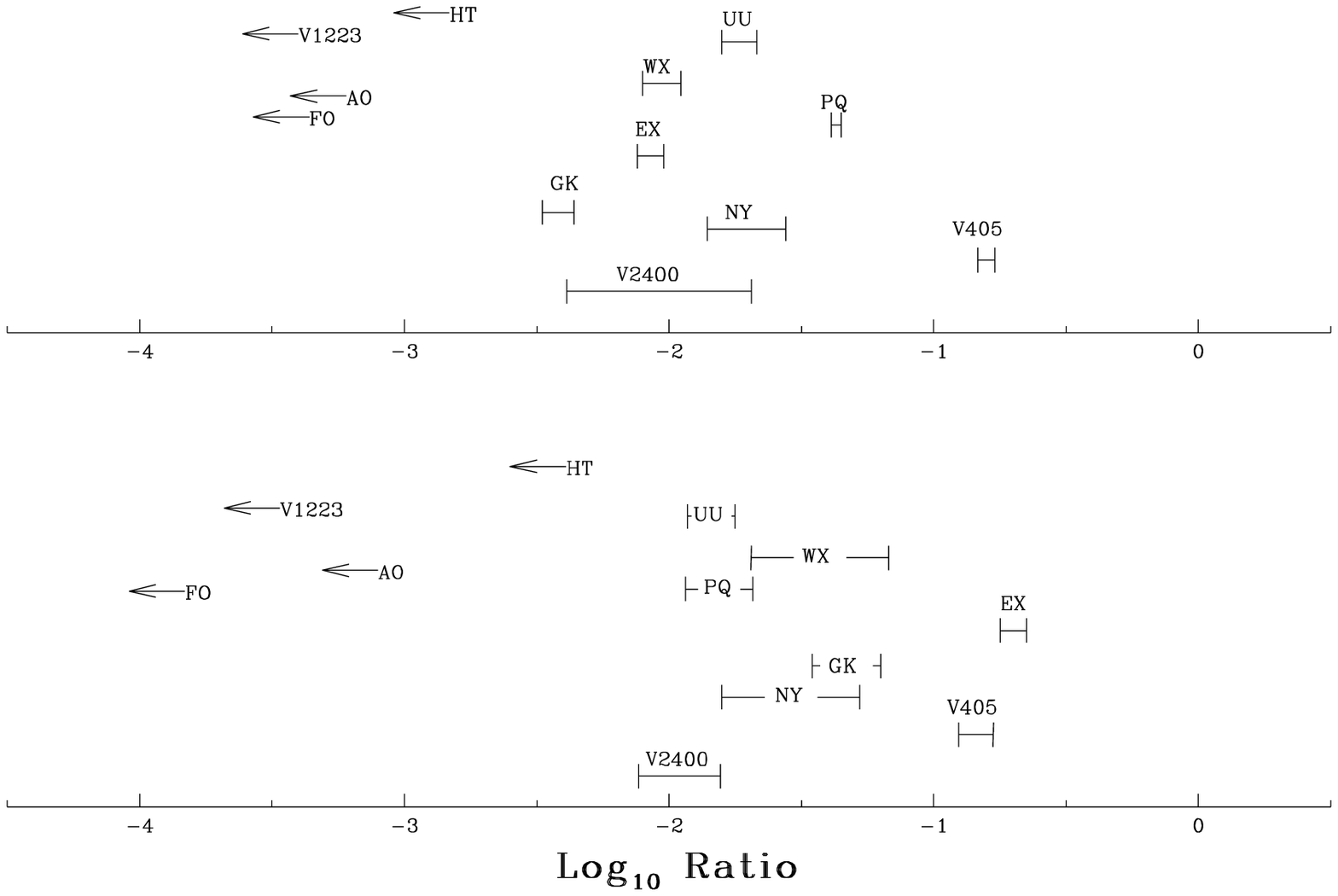}
\caption{Softness ratios of the IPs observed with \xmm, defined as
$F_s/4F_h$,
where $F_s$ and $F_h$ are the fluxes of the soft blackbody and hard plasma
components respectively, calculated over the 0.2--12 keV energy range
covered
by \xmm. \emph{Upper panel}: Ratios calculated from the spectral fits.
\emph{Lower panel}: The ratios calculated from the spectral fits, after the
absorption components were removed.}
\label{fig:scatterlim}
\end{figure*}

\begin{figure*}
\plotone{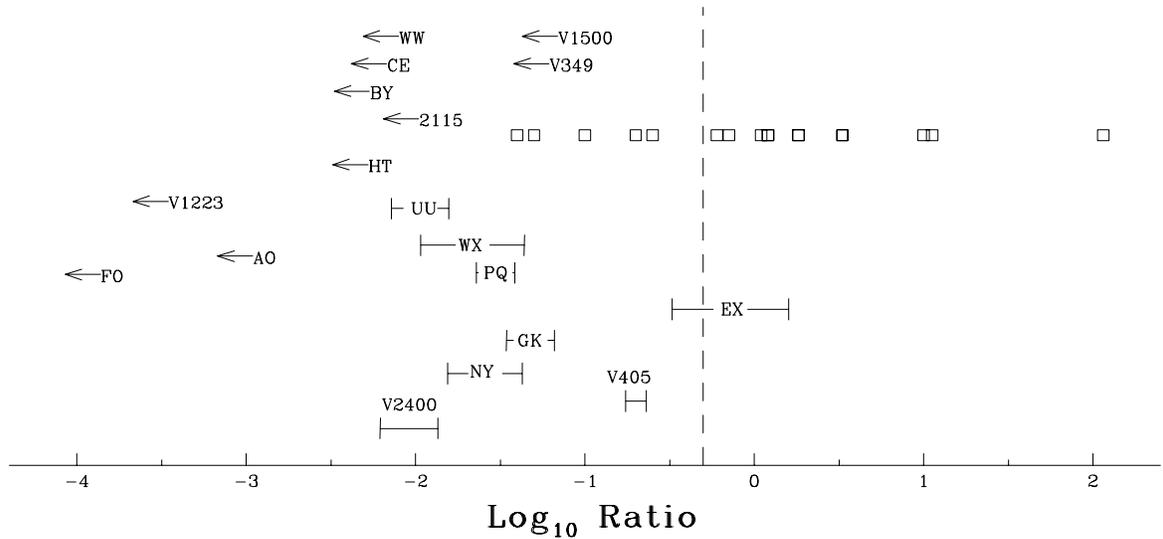}
\caption{As for Fig.~\ref{fig:scatterlim}, but with the effects of absorption
removed and $F_s$ and $F_h$ extended over all energies. The ratios for the
polars given in RC04 are also shown (hollow squares); RC04 did not quote errors.
The uppermost six systems are the polars in which RC04 found no blackbody
component. We have found an upper limit for these systems as we did for the IPs.
The dashed line corresponds to a softness ratio of 0.5; systems with a higher
softness ratio exhibit a `soft excess'.}
\label{fig:scatterbol}
\end{figure*}

The softness ratios are shown in Figs.~\ref{fig:scatterlim}
and~\ref{fig:scatterbol}. We show the observed ratio, the ratio of unabsorbed
fluxes over the 0.2--12 keV range, and the ratio of unabsorbed fluxes calculated
over all energies. These bolometric fluxes and softness ratios are given in
Table~\ref{tab:soft}. For the systems with no detectable soft component we show
the upper limit calculated for a 60-eV blackbody, and present the fluxes and
ratios for a range of blackbody temperatures in Table~\ref{tab:hard}.

%%%%%%%%%%%%%%%%%%%%%%%%%%%% SOFT IPS %%%%%%%%%%%%%%%%%%%%%%%%%%%%%%

\begin{deluxetable*}{llll}
\tablecaption{The unabsorbed, bolometric fluxes of the soft and hard
components for
those systems which show blackbody emission.}
\tablehead{
\colhead{Object} & \colhead{$F_{h, \rm bol}$} & \colhead{$F_{s, \rm bol}$} &
\colhead{Ratio} \\
& \colhead{(erg s$^{-1}$ \cms)} & \colhead{(erg s$^{-1}$ \cms)} }
\startdata
V405~Aur   &  $5.1\tim{-11}$ (+3.6,  $-$1.1)    & $4.3\tim{-11}$ (+2.4,
$-$1.2)    & 0.211 (+0.018, -0.038) \\
GK~Per     &  $1.20\tim{-9}$ (+0.25, $-$0.06)   & $2.29\tim{-10}$
(+0.95, $-$0.62) & $4.8\tim{-2}$ (+1.8, $-$1.3) \\
NY~Lup     &  $4.15\tim{-11}$ (+12.7, $-$0.18)  & $4.3\tim{-12}$ (+9.9,
 $-$1.4)   & $2.6\tim{-2}$ (+1.6,  $-$1.1) \\
V2400~Oph  &  $9.2\tim{-11}$  (+4.1,  $-$1.9)   & $3.3\tim{-12}$ (+2.1,
$-$1.5)    & $8.9\tim{-3}$ (+4.6, $-$2.8) \\
PQ~Gem     &  $1.07\tim{-10}$ (+0.40, $-$0.23)  & $1.33\tim{-11}$
(+0.13, $-$0.17) & $3.11\tim{-2}$ (+0.74, $-$0.83) \\
EX Hya     &  $3.95\tim{-10}$ (+0.26, $-$0.20)  & $1.59\tim{-10}$
(+3.47, $-$0.62) & 1.00 (+0.48, $-$0.20) \\
WX Pyx     &  $7.51\tim{-12}$ (+0.35, $-$0.93)  & $6.0\tim{-13}$ (+5.9,
$-$2.9)    & $2.00\tim{-2}$ (+2.38, $-$0.93)\\
UU Col     &  $6.81\tim{-12}$ (+2.29, $-$0.80)  & $3.04\tim{-13}$
($\pm$0.98) & $1.12\tim{-2}$ ($\pm$0.45) \\
\enddata
\tablecomments{The ratio is defined as in Fig.~\ref{fig:scatterlim}.
Errors are given
to the same power of ten as the values.}
\label{tab:soft}
\end{deluxetable*}

%%%%%%%%%%%%%%%%%%%%%%%%%%%% HARD IPS %%%%%%%%%%%%%%%%%%%%%%%%%%%%%%

\begin{deluxetable*}{llccc}
\tablecaption{The unabsorbed, bolometric fluxes from the systems with no
detectable
soft X-ray component, and the upper limit of the softness ratio, for a
range of
temperatures.}
\tablehead{ \colhead{Object} & \colhead{$F_{h, \rm bol}$} &
\colhead{Ratio$_{40 eV}$}
& \colhead{Ratio$_{60 eV}$} & \colhead{Ratio$_{80 eV}$} \\
&  &  \colhead{(erg s$^{-1}$ \cms)}}
\startdata
FO~Aqr     & $2.71\tim{-10}$(+0.65, $-$0.18)   & $<4.6\tim{-4}$  &
$<1.4\tim{-4}$  & $<8.0\tim{-5}$  \\
AO~Psc     & $1.51\tim{-10}$(+0.09, $-$0.11)   & $<4.3\tim{-3}$  &
$<1.1\tim{-3}$  & $<6.2\tim{-4}$  \\
HT~Cam     & $8.48\tim{-12}$(+0.53, $-$0.38)   & $<2.5\tim{-2}$  &
$<5.3\tim{-3}$  & $<2.6\tim{-3}$  \\
V1223~Sgr  & $2.96\tim{-10}$( $\pm0.13$)       & $<8.5\tim{-4}$  &
$<3.5\tim{-4}$  & $<2.3\tim{-4}$  \\
\enddata
\label{tab:hard*}
\end{deluxetable*}

\section{Results}
\label{sec:results}

\begin{figure*}
\plotone{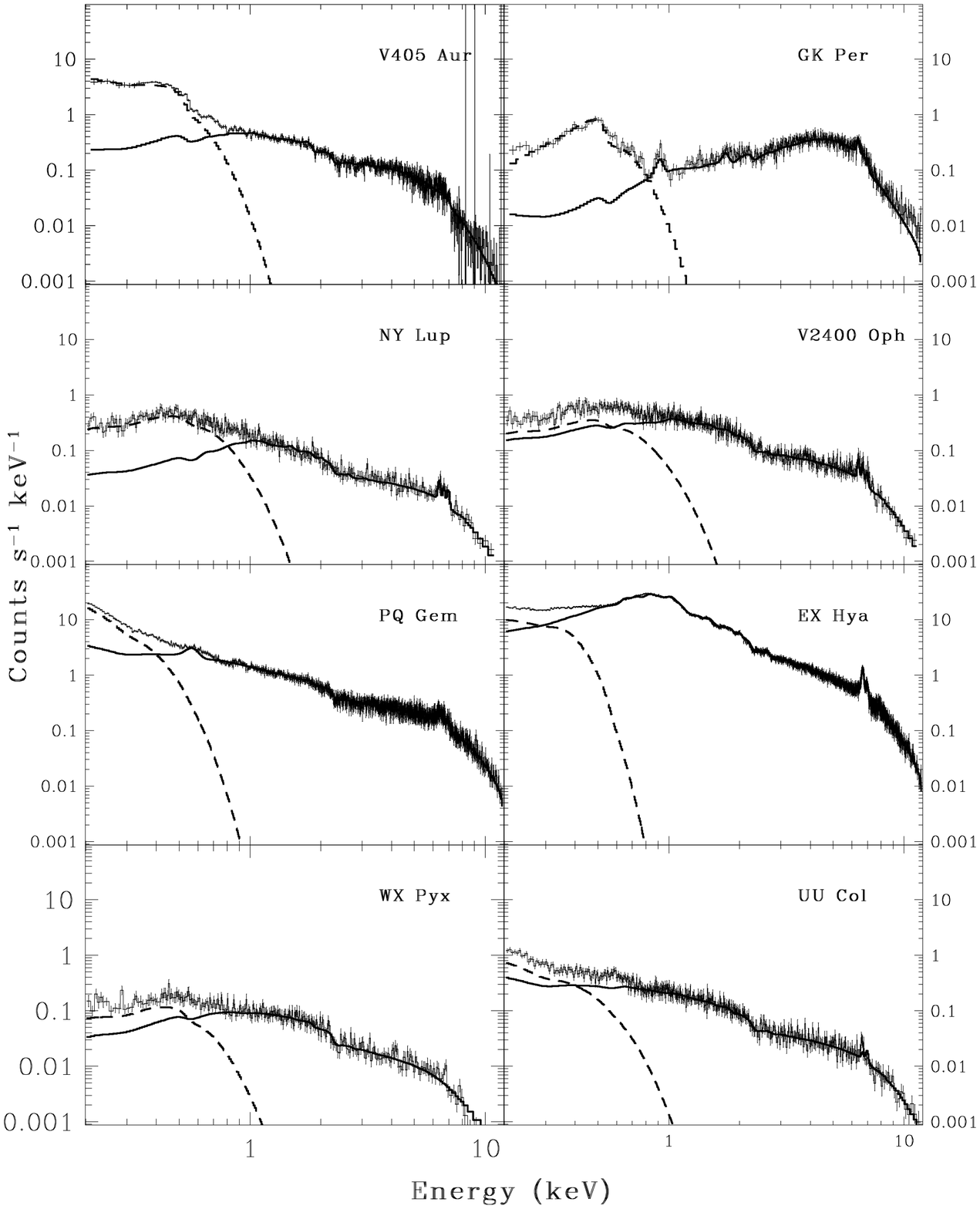}
\caption{The EPIC-pn spectra of the eight IPs for which the best-fitting models
contain a blackbody component. The solid line shows the hard component; the
broken line the blackbody. For V405~Aur we have shown the MOS-1 spectrum, since
the pn camera did not collect any data.}
\label{fig:soft}
\end{figure*}

\begin{figure*}
\plotone{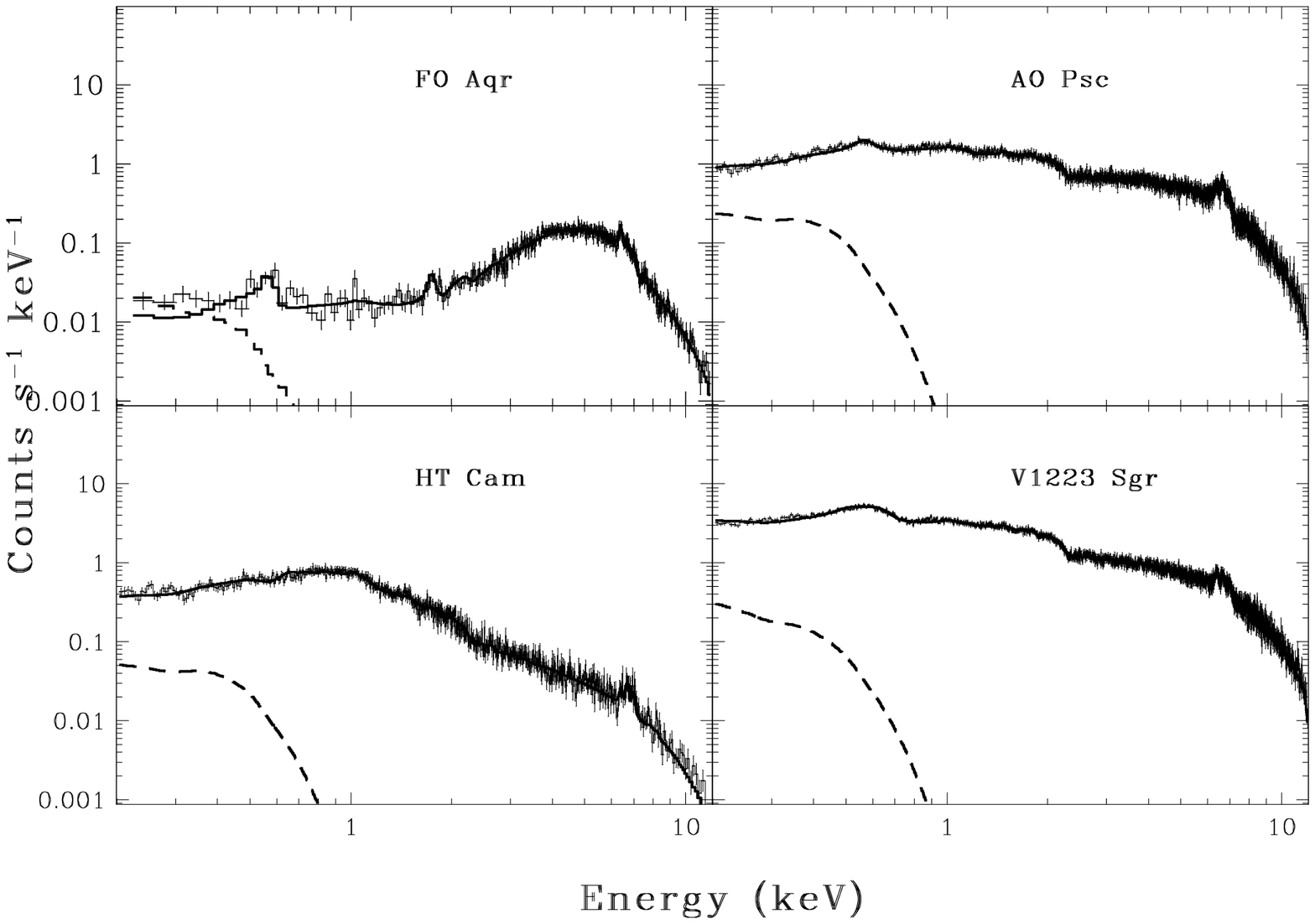}
\caption{The EPIC-pn spectra of the four IPs for which the best-fitting
model does not contain a blackbody component. The solid line shows the
best-fitting model. The broken line shows the upper limit to a blackbody component,
given a temperature of 60 eV. For FO~Aqr we show the MOS-1 data, as the
signal-to-noise ratio of the pn data is worse.}
\label{fig:hard}
\end{figure*}

We show the spectra for the systems with a blackbody component in
Fig.~\ref{fig:soft}, and for those without in Fig.~\ref{fig:hard}. For the
latter we have also shown the upper limit determined for a 60-eV blackbody
component.

For FO~Aqr, AO~Psc, V1223~Sgr and HT~Cam we found no evidence for a soft
component, in agreement with previous observations, (see
\citet{Norton92,Hellier96,Beardmore00,Evans05a} respectively).

\subsection{V405~Aur}
\label{sec:v405}

The \xmm\ observation of V405~Aur contains systematic discrepancies between the
two EPIC-MOS instruments below 0.4 keV. However, when processed under {\sc sas}
7.0 these are at a much lower level than when \citet{Evans04a} analysed the
data, and we have made no allowance for these discrepancies in the fit. Note
also that as there is no pn data for V405~Aur, we have no estimate of the
effects of systematics discussed in Section~\ref{sec:obs}, so our errors are
likely to be underestimates.

The best-fitting blackbody temperature was $kT=64.78^{+0.81}_{-1.11}$ eV. This
is significantly higher than the $40\pm4$ eV reported by \citet{Evans04a}
analysing the same observation, however they used two \mekal s to fit the hard
component whereas we used the stratified column model. Since the calibration has
also changed since \citet{Evans04a}, we analysed our better-calibrated data
using their model, and found a fit in agreement with theirs. This demonstrates
that the results are somewhat model dependent; the stratified column model is
likely to be the more physically realistic. Fitting the hard component with a
single, high temperature plasma, \citet{Haberl94}  found a blackbody temperature
of 49--64 eV (from \emph{ROSAT\/} data) and \citet{deMartino04} found $73\pm14$
eV (using \emph{BeppoSAX}).

Our fitted hydrogen column of $3.46^{+0.41}_{-0.31}\tim{20}$\ \cms\ agrees with
that of \citet{deMartino04} [$(4\pm2)$\tim{20} \cms] but not with those of
\citet{Haberl94} or \citet{Evans04a} who reported $(5.7\pm0.3)\tim{20}$ and
$(10.6^{+0.9}_{-1.2})\tim{20}$\ \cms\ respectively. However, the fitted column
will depend on the emission model used, so some discrepancy is expected.

\subsection{GK~Per}
\label{sec:gkper}

A soft blackbody component was necessary to model the \xmm\ spectrum of GK~Per
as previously found by  \citet{Vrielmann05}. They reported a blackbody
temperature of $59.6\pm0.2$ eV absorbed by a column of $(3.2\pm 0.2)\tim{21}$
\cms. Our temperature of $62\pm2$ eV and column of $(2.3\pm0.2)$ \cms\ are very
similar, though not formally in agreement. Note that \citet{Vrielmann05}
parameterised the hard emission using a bremsstrahlung component and a \mekal,
supporting our assertion above that these results are model dependent.

\subsection{NY~Lup}
\label{sec:rxj}

\citet{Haberl02} analysed this \xmm\ observation of NY~Lup (=RX\,J154814) and
found a soft component with a blackbody temperature of 84--97 eV and a column
density of (11.7--15.5) \tim{20} \cms. Our values of $kT_{\rm
bb}=104^{+21}_{-23}$ eV and $n_H=(7.8\pm3.9)\tim{21}$ \cms\ agree.

\subsection{V2400~Oph}
\label{sec:v2400}

V2400~Oph was identified as a soft IP by \citet{deMartino04}, who analysed a
\emph{BeppoSAX\/} observation and reported a blackbody temperature of $103\pm10$
eV and absorption column $(46^{+12}_{-13})$\tim{20} \cms. We find a blackbody
temperature of $117^{+33}_{-44}$ eV, in agreement with this result, but a
slightly lower absorption column of $(7.0^{+2.9}_{-4.9})\tim{20}$ \cms. This is
probably because de~Martino used a single \mekal\ and a single partial-covering
absorber to model the hard emission, whereas we used the stratified column model
and two partial covering absorbers.

\subsection{PQ~Gem}
\label{sec:pqgem}

PQ~Gem was the first IP found to have a soft-X-ray component \citep{Mason92}.
This component was also present in the \xmm\ data, with a best-fitting blackbody
temperature of $47.6^{+2.9}_{-1.4}$ eV, in agreement with the $46^{+12}_{-23}$
eV of \citet{Duck94} from \emph{ROSAT\/} data, and $56^{+12}_{-14}$ eV of
\citet{deMartino04} from \emph{BeppoSAX\/} data. The fitted column density goes
to zero, which is likely to be an artefact of fitting a complex absorption with
too simple a model. We quote an upper limit of $3\tim{19}$ \cms\ based on the
phase-resolved modelling of \citet{Evans06}.

\subsection{EX~Hya}
\label{sec:exhya}

The best-fitting model for EX~Hya used a blackbody component, which has not been
previously reported in this system.  However even with this component, the
procedure outlined in Section~\ref{sec:obs} resulted in a poor fit
(\rchisq=2.1). A possible reason for this is our choice of absorption model. We
have used a cold absorber in our models since the data do not warrant the extra
parameters in ionised absorption models, even though one expects any absorbing
material (e.g.\ the accretion curtains) to be ionised. We therefore tried
various ionised absorption models, but gained only a minor improvement to the
fit. We thus reverted to the cold absorber model for consistency with the rest
of this paper. We also tried using phase-resolved spectroscopy, in case the poor
fit was the result of averaging phase-variant parameters, however this still did
not yield an acceptable fit. We have nonethless included our results for EX~Hya,
for completeness, but due to the poor fit, we do not much place much
weight on the EX~Hya data when considering our results.

As the distance to EX~Hya is known (64.5$\pm1.2$ pc: \citet{Beuermann03}), we
can determine the size of the accretion footprint from the soft X-ray flux.
Table~\ref{tab:soft} gives this as $(1.59^{+3.47}_{-0.62})\tim{-10}$ ergs \cms\
s$^{-1}$  with a temperature of $31.0^{+1.3}_{-2.4}$ eV, from which we compute
an emitting area $(8.4^{+29.8}_{-4.2})\tim{13}$ cm$^{2}$. \citet{Suleimanov05}
gave the mass of the white dwarf in EX~Hya as $0.5\pm0.05$ \msol, thus the
observed blackbody emitting area in EX~Hya covers $(7.3^{+29.3}_{-4.0})\tim{-4}$
of the white-dwarf surface.

\subsection{UU~Col}
\label{sec:uucol}

UU~Col was identified as a soft IP by \citet{Burwitz96}. \citet{deMartino06}
have recently confirmed this with a detailed analysis of the \xmm\ observation. 
They reported a blackbody temperature of $49.7^{+5.6}_{-2.9}$ eV, which is lower
than our value of $73^{+20}_{-9}$ eV, however in their model the blackbody is
absorbed by the partial covering absorber, and no simple absorber is present.

\subsection{WX~Pyx}
\label{sec:wxpyx}

The XMM observation of WX Pyx, the only X-ray observation of this star to date,
has a relatively low statistical quality.  It was previously analysed by
Schlegel (2005) who did not report looking for a blackbody component.   However,
we find that adding a blackbody does  significantly improve the fit.

\subsection{Comparison with the polars}
\label{sec:polars}

In Fig.~\ref{fig:scatterbol} we have plotted the softness ratios of both the IPs
and the polars (from RC04). For the polars which RC04 reported not to have a
blackbody, we obtained the spectra as extracted an calibrated by RC04 (Ramsay,
private communication), and fitted them in the same way as the IPs 
(Section~\ref{sec:obs}) to obtain an upper limit. 

The chief difference in the two distributions is that while several polars show
a softness ratio $>0.5$, no IP can be confirmed to do this, and it can be
excluded for all but EX~Hya -- for which our results are uncertain
(Section~\ref{sec:exhya}). The `soft excess' in polars is believed to arise due
to `blobby accretion' (e.g.\ \citet{Kuijpers82}). In this model, dense blobs of
matter penetrate into the white dwarf photosphere and the energy is thermalised
to a blackbody.

Whether such accretion occurs in IPs has not been widely discussed in the
literature. \citet{Hellier02} suggested that viscous interactions in
an accretion disc would destroy blobs, although \citet{Vrielmann05}
interpreted flares in the lightcurve of GK~Per as resulting from the
accretion
of blobs.  Our findings suggest that blobby accretion is not significant in
IPs.

\section{Discussion}
\label{sec:disc}

The `polar' class of magnetic cataclysmic variable has long been known to be
characterised by a soft blackbody component (e.g.\ \citet{King87}). This is
considered to arise from the white-dwarf surface, heated either by reprocessing
of hard X-rays from the accretion column, or by thermalisation of blobs of
accretion (e.g.\ \citet{Kuijpers82}). In contrast, IPs were thought to lack this
component (e.g.\ \citet{King90}). However, observations with {\sl ROSAT\/} found
a blackbody component in some IPs, leading \citet{Haberl95} to suggest that
there were two spectrally distinct classes of IP. This raised the question of
why.

To address this we have conducted a systematic survey of the spectral
characteristics of the IPs observed with \xmmn, which has much greater spectral
coverage and throughput than \emph{ROSAT}.

We find that, of twelve IPs analysed, eight show a soft blackbody component
while four do not. This suggests that a blackbody is a normal component of IPs,
and hence of accretion onto magnetic white dwarfs, and that the spectra differ
only in degree.

We thus ask what causes the differing visibility of the soft component. There
does not appear to be any correlation with the white-dwarf mass [see
\citet{Cropper99,Ezuka99,Ramsay00,Suleimanov05} for mass estimates], nor any
obvious correlation with the orbital period.

In polars, systems with higher magnetic field strengths appear to have higher
softness ratios (e.g.\ \citet{Ramsay94}). Of the IPs in our sample showing
polarisation, and thus known to have a relatively strong field (5--20 MG), all
(PQ~Gem, V405~Aur and V2400 Oph) show a blackbody component, while the four
stars showing no blackbody emission (FO Aqr, AO~Psc, HT~Cam \&\ V1223~Sgr) do
not show polarisation. We give a possible explanation for this after discussing
the role of absorption.

We first consider the simple absorber, which is probably of interstellar origin.
The detection of a blackbody component in most systems shows that interstellar
absorption is not sufficient to extinguish the soft emission. Further, the
systems with no detected blackbody component do not have higher interstellar
columns than those with a blackbody (Table~\ref{tab:comps}), so this absorption
cannot explain the differing visibility of the soft component.

We thus turn to the partial-covering absorption, which in IPs is predominantly
caused by the accretion curtains crossing the line of sight. Here we find that
the systems where the lightcurves are dominated by deep absorption dips owing to
the accretion curtains (FO~Aqr, V1223~Sgr and AO~Psc; see
\citet{Beardmore98,Beardmore00,Hellier91} respectively) do tend to be those
which lack a blackbody component. In contrast, systems showing a blackbody
component, such as V405~Aur, NY~Lup, EX~Hya and V2400 Oph, tend to be systems
where the lightcurves suggest that the accretion curtains do not hide the
accretion footprints (see \citet{Evans04a,Haberl02,Allan98,Hellier02}
respectively).

\begin{figure}
\plotone{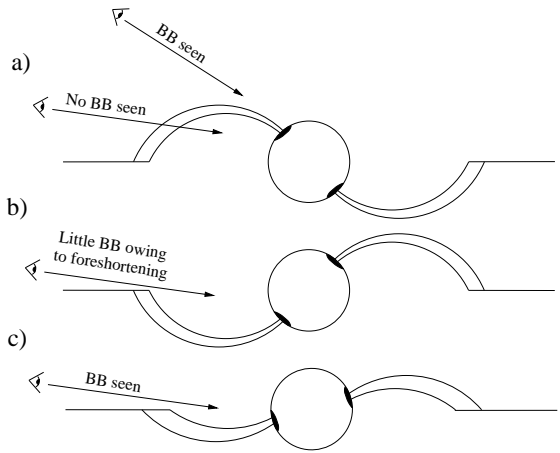}
\caption{The factors that affect blackbody emission in an IP.
\newline a) When the upper magnetic pole is on the visible face, blackbody
emission will only be seen if the inclination is such that the
heated accretion region is visible above the accretion curtains.
\newline b) When the lower pole is on the visible face, it will
likely be too foreshortened for us to detect blackbody emission.
\newline c) In UU~Col the magnetic axis is highly inclined, so the
foreshortening seen in b) is reduced and blackbody emission is seen.}
\label{fig:schem}
\end{figure}

We thus suggest that the major reason why some IPs don't show a blackbody
component is simply that the heated region near the accretion footprint is
hidden by the accretion  curtains, while in other IPs it is not, the difference
being the result of the system inclination and  the magnetic colatitude (see
Fig.~\ref{fig:schem}). Coupled with this is the effect of foreshortening, such
that an optically thick heated region will not produce much blackbody emission
if it is only seen while on the white-dwarf limb, rather than in the middle of
the face.

A proper investigation of this idea would need knowledge of the size and
location of the accretion footprints and of the surrounding heated polecaps, so
that we could estimate the difference absorbing columns of different spectral
components, and how these vary with spin-cycle phase. However, this information
is not known for the majority of IPs. The softness ratio might conceivably also
vary with parameters such as accretion rate and white-dwarf mass, which are
again only poorly known.

However, as a test of our ideas, we can outline how they might apply to the
remaining systems in our sample which we did not consider when forming the
model, namely HT~Cam, GK~Per, PQ~Gem and UU~Col.

In PQ~Gem the accretion curtains do cause an absorption dip when they obscure
the accretion footprints.  However, the geometry of this star is relatively well
determined \citep{Potter97,Mason97,Evans06} and it appears that the heated
polecap is grazingly visible above the accretion curtain for part of the cycle;
thus it shows both an absorption dip and a soft blackbody, and is on the
boundary between the two cases illustrated in the upper panel of Fig.~5.

UU~Col also shows an absorption dip when the accretion curtains obscure the
upper pole, and also shows blackbody emission. \citet{deMartino06} proposed that
the blackbody emission comes from the lower pole, viewed when that pole is
closest to us (lowest panel of Fig.~5). We thus suggest that UU~Col has an
abnormally high inclination of the magnetic dipole, such that the lower pole is
not foreshortened as much as in other IPs where no blackbody component is seen.
V405~Aur is another system that appears to have a highly inclined dipole, such
that blackbody emission from the lower pole is significant, leading in that
system to a double-peaked soft-X-ray lightcurve \citep{Evans04a}.

\begin{figure}
\plotone{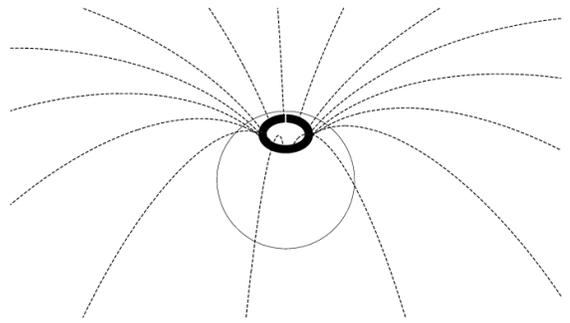}
\caption{Schematic diagram of GK~Per in outburst. Accretion occurs from all
azimuths, resulting in a circular blackbody-emitting region (dark ring).
As can
be seen, even when the accretion curtains lie across our line-of-sight,
part of
this region is unobscured.}
\label{fig:gk}
\end{figure}

In contrast to all the other IPs, the \xmm\ data of GK~Per reported here were
collected during an outburst. \citet{Hellier04} have argued that during outburst
the accretion occurs from all azimuths, forming a complete accretion ring at the
poles. As illustrated in Fig.~\ref{fig:gk}, this means that some portion of the
heated polecap is likely to be visible  `behind' the magnetic pole, where
accretion does not normally occur. Thus in GK~Per in outburst we see a system
with both strongly absorbed X-ray emission (from in front of the magnetic pole)
and a blackbody component.

Lastly, we consider HT~Cam. This shows very little sign of absorption, and its
lightcurve can explained without any absorption effects \citep{Evans06}. Yet it
shows no blackbody emission, in apparent contradiction to our model. However, as
previously suggested by \citet{deMartino06} and \citet{Evans06}, it appears that
HT~Cam has an exceptionally low accretion rate (partly accounting for the lack
of absorption). If so, it could be that the blackbody component is simply too
cool to be detected in the \xmm\  bandpass. We note that the blackbody
temperature in EX~Hya, the other star in our sample below the period gap, is
lower than in the others (Table~\ref{tab:comps}), and that in HT~Cam might be
lower still.  

\section{Summary}

We have analysed data from \xmm\ observations of 12 intermediate polars and find
that a soft blackbody component is a common feature of their X-ray spectra. We
suggest that in the systems showing no blackbody emission the heated accretion
polecaps are largely hidden by the accretion curtains, or are only visible when
on the white dwarf limb and highly foreshortened. Thus IPs with lightcurves
dominated by absorption dips caused by the passage of accretion curtains across
the line of sight tend to show no blackbody emission. Further, these are also
the systems least likely to show polarisation, since the cyclotron-emitting
column will also be obscured by the accretion curtains, or would be beamed away
from us if the accretion region were on the white-dwarf limb. After comparing
the blackbody emission seen in IPs with that seen in polars, we conclude that
the blobby emission responsible for soft X-ray excesses in polars does not occur
in IPs.

\section*{Acknowledgements}

We thank Gavin Ramsay for providing us with the spectra of the polars with no
detectable soft component.

{\it Facilities:} \facility{XMM ()}

\end{document}